\documentclass[footinbib,aps,pra,reprint,superscriptaddress,nopacsm]{revtex4-1}

\usepackage{graphicx}
\usepackage{amsmath,amsfonts,amssymb,bbm,bm}
\usepackage{verbatim}
\usepackage{braket}
\usepackage{mathtools}
\usepackage[absolute]{textpos}
\usepackage{ragged2e}
\usepackage[hidelinks]{hyperref}
\usepackage[capitalize]{cleveref}
\crefname{section}{Sec.}{}

\usepackage{xcolor}

\begin{document}
\title{Characterization of Quantum Frequency Processors}

\author{Hsuan-Hao Lu}
\email{luh2@ornl.gov}
\affiliation{Quantum Information Science Section, Oak Ridge National Laboratory, Oak Ridge, Tennessee 37831}
\author{Nicholas A. Peters}
\affiliation{Quantum Information Science Section, Oak Ridge National Laboratory, Oak Ridge, Tennessee 37831}
\author{Andrew M. Weiner}
\affiliation{School of Electrical and Computer Engineering and Purdue Quantum Science and Engineering Institute, Purdue University, West Lafayette, Indiana 47907, USA}
\author{Joseph M. Lukens}
\email{joseph.lukens@asu.edu}
\affiliation{Quantum Information Science Section, Oak Ridge National Laboratory, Oak Ridge, Tennessee 37831}
\affiliation{Research Technology Office and Quantum Collaborative, Arizona State University, Tempe, Arizona 85287, USA}
\date{\today}

\begin{abstract}
Frequency-bin qubits possess unique synergies with wavelength-multiplexed lightwave communications, suggesting valuable opportunities for quantum networking with the existing fiber-optic infrastructure. Although the coherent manipulation of frequency-bin states requires highly controllable multi-spectral-mode interference, the quantum frequency processor (QFP) provides a scalable path for gate synthesis leveraging standard telecom components. Here we summarize the state of the art in experimental QFP characterization. Distinguishing between physically motivated ``open box'' approaches that treat the QFP as a multiport interferometer, and ``black box'' approaches that view the QFP as a general quantum operation, we highlight the assumptions and results of multiple techniques, including quantum process tomography of a tunable beamsplitter---to our knowledge the first full process tomography of any frequency-bin operation. Our findings should inform future characterization efforts as the QFP increasingly moves beyond proof-of-principle tabletop demonstrations toward integrated devices and deployed quantum networking experiments.
\end{abstract}

\maketitle

\section{Introduction}
\label{sec:intro}
In modern lightwave communications, the optical and electronic domains work in concert for the distribution and processing of information~\cite{Agrell2016}. Light fields carry data streams with dense multiplexing down low-loss optical fiber, while electronic circuits perform digital processing for logical computations, error correction, and packet switching. The inherently ultrawide bandwidth available to photonic systems---along with the potential for greater speed and efficiency by avoiding optical-to-electrical conversion---has inspired a plethora of research into all-optical processing approaches challenging this status quo and aiming to supplant electronic digital logic with all-optical interactions~\cite{Willner2014}. 
And in the case of photonic \emph{quantum} information, all-optical processing is more than just a pathway for improved performance: instead, it is a prerequisite, since conversion to the digital domain represents a measurement that irreversibly collapses the quantum state.

Remarkably, Knill, Laflamme, and Milburn (KLM) showed in 2001 that the photonic interactions necessary for universal quantum information processing can be realized with multiphoton interference in linear-optical circuits, combined with detection and feed-forward~\cite{Knill2001}. Although experimentally challenging, the KLM scheme is in principle scalable and has stimulated a large body of research on linear-optical quantum computing~\cite{Kok2007}, including important developments in measurement-based computing approaches---both discrete-~\cite{Nielsen2004} and continuous-variable~\cite{Menicucci2006}---that appear promising for a large-scale photonic quantum computer~\cite{Rudolph2017}. Beyond computing alone, KLM concepts form the foundation for all-photonic quantum repeater designs as well, which could eliminate the need for quantum memories and two-way communications~\cite{Azuma2015, Munro2015, Pant2017}.
Although originally proposed for spatio-polarization encodings, KLM approaches have been extended into a variety of nontraditional degrees of freedom, including time bins~\cite{Humphreys2013} and pulsed modes~\cite{Brecht2015}. 

Mandatory to any KLM-like scheme is the ability to construct arbitrary unitary transformations. In the case of frequency-bin encoding---intriguing for its compatibility with wavelength-multiplexing in optical fiber, on-chip photon sources, and frequency-disparate quantum interconnects---it was unclear whether arbitrary unitaries could be synthesized with standard fiber-optic components, until the quantum frequency processor (QFP) was introduced and analyzed in 2017~\cite{Lukens2017}. Based on an alternating series of electro-optic phase modulators (EOMs) and Fourier-transform pulse shapers, the QFP realizes frequency-bin transformations through what can conceptually be viewed as successive phase-only filters in the time and frequency domains. In \cite{Lukens2017}, explicit designs for a universal gate set (single-qubit phase and Hadamard $H$, and two-qubit controlled-$Z$) were discovered, supported further by an argument for linear scaling of resources (EOMs and pulse shapers) with the number of modes. In the intervening years, many experiments have been performed demonstrating basic QFP gates, on both single-~\cite{Lu2018a, Lu2020b, Lu2022b} and two-photon states~\cite{Lu2018b, Lu2019a, Lingaraju2022}. As with any experimental demonstration, it has been critical to confirm the degree of agreement between the realized QFP operation and that expected from theory, for which a variety of probes have been enlisted, including bright frequency combs, weak coherent states, and entangled photons. 

In this article, we present an introduction and overview of the techniques currently available for QFP characterization, 
complementing previous high-level reviews~\cite{Kues2019, Lu2019c} with a much more thorough focus on the underlying assumptions and information provided by various characterization approaches. Following background on the theory and implementation in Sec.~\ref{sec:background}, Secs.~\ref{sec:OpenBox} and \ref{sec:BlackBox} summarize QFP characterization experiments organized in terms of successively fewer assumptions, including an entirely new experiment performing Bayesian quantum process tomography (QPT) of the tunable beamsplitter---the first QPT characterization of a frequency-bin gate. Throughout, we find it useful to frame the discussion around ``open box'' and ``black box'' techniques, distinguished by their treatment of the QFP's inner workings. In addition to furnishing an organized introduction to this important subfield of time-frequency quantum information, our QPT results open the door for impartial characterization of emerging QFP designs---e.g., photonic integrated circuits---comprised of EOMs and pulse shapers whose properties may deviate strongly from ideal models previously assumed for discrete devices.

\section{QFP Background}
\label{sec:background}

\begin{figure}[tb!]
\centering
\includegraphics[width=3.5in]{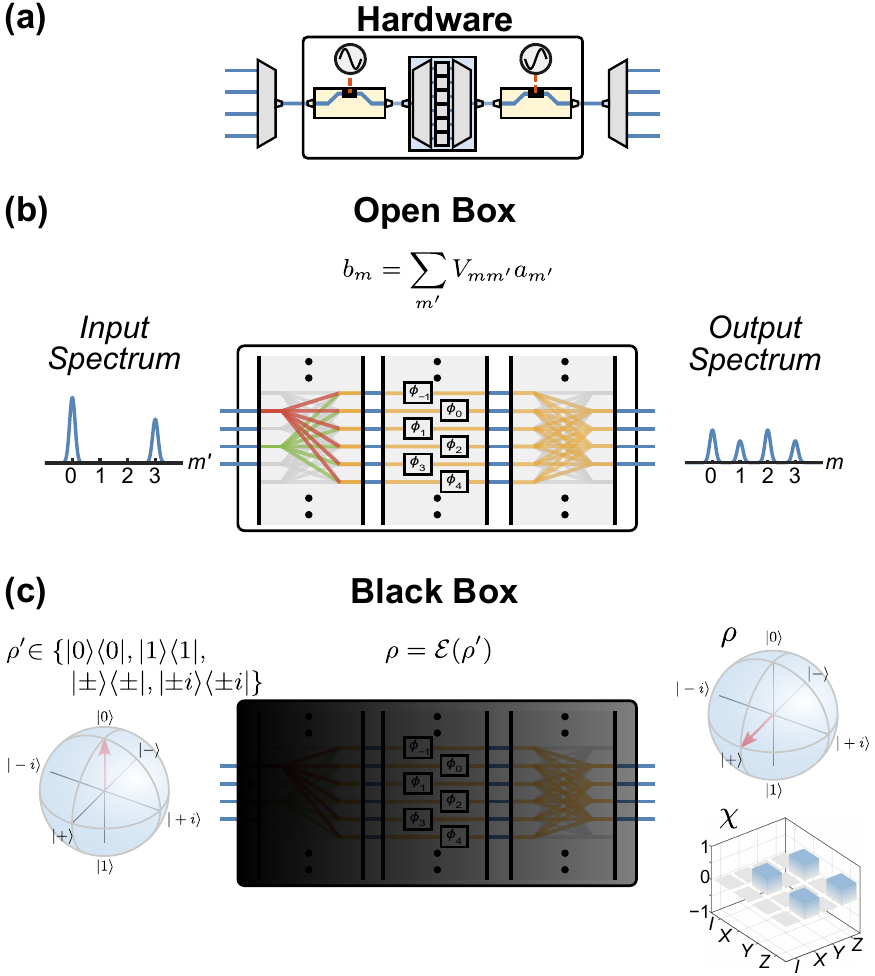}
\caption{QFP characterization concepts. (a)~Physically, a QFP consists of an alternating series of EOMs and pulse shapers, here shown in a three-element instantiation (EOM/pulse shaper/EOM) flanked by spectral multiplexers. (b)~In the open box viewpoint, the output frequency-bin operators are assumed to follow from the input through a linear transformation $V$, which can be characterized by measuring the output spectrum of superposition probe states. (c)~In the black box viewpoint, the QFP is treated as a generic quantum channel to be analyzed by quantum state or process tomography, returning output state $\rho$ or process matrix $\chi$, respectively. The examples in (b) match the mode transformation of a Bell state analyzer~\cite{Lingaraju2022}, while those in (c) correspond to an ideal Hadamard gate~\cite{Lu2018a}.}
\label{fig:1}
\end{figure}

\subsection{Theory}
\label{sec:theory}
As noted above, the distinction between open box and black box characterization provides a useful framework through which to view previous and current QFP experiments. The notion of a black box in engineering contexts is simple to define: a device to which one has access only to the inputs and outputs, making the system opaque from the perspective of the outside world. The opposite of such a system has been dubbed, e.g., an ``open box'' or ``white box'' in the literature, defined as device for which at least some of the internals are known to the tester; in computer science, both open~\cite{Ostrand2002a} and black box~\cite{Ostrand2002b} software testing are common. Yet while black box concepts have a long history in quantum information processing---indeed the original proposal for QPT used the term explicitly~\cite{Chuang1997}---open box terminology has not taken hold in this field, at least in our experience. Nevertheless, for the QFP the distinction seems particularly apt, given the gradual increase in complexity of characterization approaches which have been demonstrated over the past few years. As will be defined mathematically below, open box QFP approaches are distinguished by assuming the QFP is a true multiport frequency-bin interferometer, per its design; black box QFP approaches make no assumptions about the QFP's insides beyond the minimum for an arbitrary quantum channel.

Fundamentally, the goal of the QFP is to controllably transform quantum states of light populating a comb of discrete frequency modes, or bins. Consider $M$ bins centered at frequencies $\omega_m=\omega_0 + m\Omega$ ($m\in\{0,1,...,M-1\}$), with $\Omega$ a fixed spacing typically in the radio-frequency (RF) domain. Each of these bins possesses an annihilation (creation) operator $a_m$ ($a_m^\dagger$) that subtracts (adds) single photons to mode $m$ when applied to any quantum state. For notational convenience, we utilize $b_m$ ($b_m^\dagger$) to denote the same operations but referenced to the output side of the quantum transformation of interest, 
so that the input (output) bins are associated with $a_m/a_m^\dagger$ ($b_m/b_m^\dagger$).

The spectral shape of these bins can in principle be of any form, provided they are bandlimited to $\omega_m\pm\frac{\Omega}{2}$. For the QFP to function as designed, these bins should possess identical lineshapes and be easily distinguished within the resolution of the pulse shapers used for line-by-line phase control~\cite{Lu2020a}. We do note that in the general quantum black box formalism, these additional assumptions are not strictly required for the QFP, but only for the system used to measure these bins at the output. Although typically based on the same technology as the QFP, the measurement system need not have the same spectral resolution as the QFP under test. Indeed, emerging QFPs based on integrated photonic circuits~\cite{Moody2022, SiPhQFP} will likely first be tested with well-characterized discrete fiber-optic components, so the distinction between QFP and measurement spectral resolutions is practically as well as theoretically significant.

Each Fock basis state---i.e., a quantum state with a specified number of photons in each mode---can be described for the input Hilbert space by a length-$M$ vector of photon occupancy numbers $\mathbf{n}' = (n_0',...,n_{M-1}')$, defined as
\begin{equation}
\label{eq:J1}
\ket{\mathbf{n}'}=\prod_{k=0}^{M-1} \frac{(a_k^\dagger)^{n_k'}}{\sqrt{n_k'!}} \ket{\mathrm{vac}},
\end{equation}
where $\ket{\mathrm{vac}}$ denotes the vacuum state over all $M$ modes. Similarly, the output is spanned by basis states $\mathbf{n} = (n_0,...,n_{M-1})$:
\begin{equation}
\label{eq:J2}
\ket{\mathbf{n}}=\prod_{k=0}^{M-1} \frac{(b_k^\dagger)^{n_k}}{\sqrt{n_k!}} \ket{\mathrm{vac}}.
\end{equation}
(As a means to aid in distinguishing between input and output, we use primed vectors $\textbf{n}'$ for the former and $\textbf{n}$ for the latter.)

The QFP transforms between these spaces via an alternating series of (i)~pulse shapers, which apply the phase filters $b_n=e^{i\phi_n}a_n$, and (ii)~EOMs, which apply a temporal phase $\varphi(t)$ periodic at frequency $\Omega$, leading to the discrete convolution $b_n=\sum_{n'} c_{n-n'}a_{n'}$  where $e^{i\varphi(t)}=\sum_k c_k e^{-ik\Omega t}$. Figure~\ref{fig:1} offers a schematic of the physical hardware for a three-element QFP, with input and output multiplexers to facilitate characterization; three-element QFPs have been the focus of experiments so far, although deeper QFP circuits remain of interest for future research. The total effect of an ideal QFP with any number of elements can be summarized as a multiport frequency-bin interferometer:
\begin{equation}
\label{eq:J3}
b_m = \sum_{m'=0}^{M-1} V_{mm'} a_{m'}.
\end{equation}
It is important to note that the Fourier series expansion of the phase modulation in general is unbounded in frequency; thus, even with the input modes restricted to a finite number $M$, a countably infinite number of output bins can be populated within and after the interferometer. Indeed, it is over this infinite set that $V$ is unitary: $\sum_{k=-\infty}^\infty V_{km}^* V_{km'} = \sum_{k=-\infty}^\infty V_{mk} V_{m'k}^* = \delta_{mm'}$. Although the intrinsically infinite-dimensional space produces interesting challenges conceptually, it can be accurately approximated by truncating to a value of $M$ sufficiently large to encompass all nonnegligible probability amplitudes excited by the input states of interest~\cite{Lukens2017}. 
In practice, the sufficiency of any truncation $M$ can be validated ex post facto by ensuring the calculated probability amplitudes decay to zero (to within some desired precision) at the edges of the domain following each EOM in the QFP circuit.

Because the QFP operates on frequency modes irrespective of the photons populating them, it can be applied to a variety of information processing scenarios, including classical lightwave communications~\cite{Lukens2020a, Lu2020a} and continuous-variable quantum frequency combs~\cite{Pizzimenti2021}, which have emerged as an exciting platform for quantum computing~\cite{Roslund2014, Pfister2020}. Nonetheless, we focus on discrete-variable encoding in this paper, the dominant paradigm in QFP characterization experiments so far. In this case, each logical basis state is associated with a specific Fock state; under dual-rail encoding, for example, a qubit is encoded by a single photon populating a superposition of two bins, one defined as $\ket{0}$ and the other as $\ket{1}$. Experimentally, we always consider the postselected regime, where all photons of the input are successfully passed through the system. Therefore any insertion loss (absent in the ideal model above) leads to a lower throughput without inherently degrading state fidelity.

In the Heisenberg picture, a fixed quantum state $\ket{\psi}$ with $N$ total photons can be represented in either the input or output basis: $\ket{\psi} = \sum_{\mathbf{n}}\beta_\mathbf{n}\ket{\mathbf{n}} =\sum_{\mathbf{n'}}\alpha_\mathbf{n'}\ket{\mathbf{n'}}$, where each sum considers only terms with $\sum_{m=0}^{M-1}n_m = \sum_{m'=0}^{M-1}n_m' = N$. Solving for $\beta_\mathbf{n}$ using Eqs.~(\ref{eq:J1}--\ref{eq:J3}) then leads to the simple vector expression $\beta_\mathbf{n} = \sum_{\mathbf{n}'}W_{\mathbf{n}\mathbf{n}'}\alpha_{\mathbf{n}'}$, where
\begin{equation}
\label{eq:J4}
W_{\mathbf{n}\mathbf{n}'} = \frac{1}{\sqrt{\prod_{p=0}^{M-1}n_p!n_p'!}} \sum_{\mathbf{k}'\in \mathrm{perms}(\mathbf{m}')} V_{m_1k_1'} \cdots V_{m_Nk_N'}.
\end{equation}
This equation corresponds to the famous ``permanents'' expression for $N$ bosons traversing an interferometer~\cite{Skaar2004, Uskov2009}. With each Fock vector $\mathbf{n}$ [$\mathbf{n'}$] 
we define a length-$N$ mode assignment vector $\mathbf{m}=(m_1,...,m_N)$ [$\mathbf{m}'=(m_1',...,m_N')$] that lists the modes occupied by each photon; the summation proceeds over all $N!$ permutations of $\mathbf{m}'$. Although this leads to repeated terms when multiple  photons populate the same mode, we have found it useful numerically, as it treats mode vectors with repeated indices the same as those without. (Removing repeated terms simply changes the factorial prefactor, as in the formulation of \cite{Skaar2004}.)

The matrix $W$ completely describes the QFP operation, and can be compared to some target $D\times D'$-dimensional matrix $T$ with elements $T_{\mathbf{n}\mathbf{n}'}$, where $D$ ($D'$) denotes the dimension of the output (input) subspace of interest. Both $D,D'\leq D_\mathrm{max}$, where $D_\mathrm{max}=\begin{psmallmatrix} N+M-1 \\ M-1 \end{psmallmatrix}$ denotes the Hilbert space dimension of $N$ photons populating $M$ modes; $D$ and $D'$ in practice are often much smaller than $D_\mathrm{max}$ since they can be limited to only the states of interest in a given problem. 
$W$ computed for the same set of states can then be compared to $T$ via matrix fidelity
\begin{equation}
\label{eq:J5}
\mathcal{F}_W = \frac{1}{\mathcal{P}_W} \left|\frac{\mathrm{Tr}~W^\dagger T}{\mathrm{Tr}~T^\dagger T} \right|^2,
\end{equation}
describing the closeness of the operation to the ideal,
and success
\begin{equation}
\label{eq:J6}
\mathcal{P}_W = \frac{\mathrm{Tr}~W^\dagger W}{\mathrm{Tr}~T^\dagger T},
\end{equation}
providing the probability that all input photons exit in the bins of interest~\cite{Uskov2009}.

At this stage, it is important to observe that the Fock state transformation matrix $W$---and hence $\mathcal{F}_W$ and $\mathcal{P}_W$---are completely determined by the mode transformation matrix $V$ [Eq.~\eqref{eq:J4}]. This assumption underpins the ``open box'' approach to QFP characterization, as summarized in Fig.~\ref{fig:1}(b); if Eq.~\eqref{eq:J3} holds, one can probe the QFP fully by sending in two-bin superposition states and measuring the output probabilities. A powerful technique first introduced in the spatial domain~\cite{Rahimi2013}, the amplitudes $|V_{mm'}|$ can be recovered from single-bin tests, while $\arg (V_{mm'})$ can be obtained by scanning the relative phase between two-bin superpositions. Significantly, even classically bright coherent states can be used, eliminating the need for single-photon detection altogether~\cite{Peruzzo2011,Spagnolo2013}. This method has been shown to be highly effective in characterizing linear-optical networks, including large-scale boson sampling circuits~\cite{Wang2019, Zhong2020}.

This open box viewpoint was adopted to characterize the first experimental QFP~\cite{Lu2018a} and has been widely deployed since, with remarkably good agreement with theory: experimental fidelities up to $\mathcal{F}_W=1 - 10^{-6}$ have been obtained for the frequency-bin Hadamard~\cite{Lu2019b}. Coupled with our understanding of the physical processes involved, there is little reason to question the validity of Eq.~\eqref{eq:J3} in most situations. Nevertheless, the theory of quantum processes permits much more general models, allowing for QFP characterization in which its inner mechanisms are completely arbitrary. 

Considering any $D'\times D'$-dimensional mixed input $\rho'$, the $D\times D$ output $\rho$ follows as~\cite{Kukulski2021}
\begin{equation}
\label{eq:J7}
\rho = \mathcal{E}(\rho') = \sum_{k=1}^{DD'} A_k \rho' A_k^\dagger,
\end{equation}
where  for trace preservation the $D\times D'$ Kraus operators resolve the identity $\sum_k A_k^\dagger A_k = \mathbbm{1}_{D'\times D'}$---a condition following from postselecting $N$ photons at the output. The open box picture represents the special case of this black box model when $A_1\propto W$ and $A_{k>1} = 0$. By preparing known inputs $\rho'$ and performing quantum state tomography (QST)~\cite{James2001} of the output $\rho$, insight into the operation $\mathcal{E}(\cdot)$ can be obtained without any reference to a multiport interferometer. For a single output $\rho$, agreement with theoretical expectations can be quantified through the state fidelity
\begin{equation}
\label{eq:J8}
\mathcal{F}_\rho = \left( \mathrm{Tr}~\sqrt{\sqrt{\rho_T} \rho \sqrt{\rho_T}} \right)^2,
\end{equation}
where $\rho_T\propto T\rho' T^\dagger$ is the output corresponding to the ideal target operation. For the entire channel, the process fidelity
\begin{equation}
\label{eq:J9}
\mathcal{F}_\Phi = \left( \mathrm{Tr}~\sqrt{\sqrt{\Phi_T} \Phi \sqrt{\Phi_T}} \right)^2
\end{equation}
serves this purpose, where the $DD'\times DD'$ density matrix $\Phi$ ($\Phi_T$) is obtained by the Choi--Jamio\l kowski isomorphism and completely specifies the measured (target) channel~\cite{Gilchrist2005}. 

Unlike the open box case, we do not define a success probability $\mathcal{P}$. For whereas the open box mapping can distinguish between output states in the targeted subspace (successes) and those outside of this space (failures), 
the black box model is intentionally oblivious to all subspaces apart from the $D$-dimensional output of interest. Considering non-trace-preserving operations could enable a meaningful success probability definition, but due to the experimental difficulty of distinguishing between failure and insertion loss ---at least without introducing modes outside of the quantum channel's purview---we concentrate on fidelity for the black box examples below.

Figure~\ref{fig:1}(c) highlights these concepts for a black box frequency-bin Hadamard. One state from a set of inputs is sent into the system and measured; in the example shown, the zero logical input $\rho'=\ket{0}\bra{0}$ produces the superposition output $\rho=\ket{+}\bra{+}$. With many such input/output tests, the full process can be reconstructed; shown is the ideal Hadamard process matrix $\chi$ expressed in the Pauli basis $(\sigma_0,\sigma_1,\sigma_2,\sigma_3) = (I,X,Y,Z)$; i.e., $\mathcal{E}(\rho') = \sum_{j=0}^3\sum_{k=0}^3 \chi_{jk} \sigma_j \rho' \sigma_k$.

To conclude this subsection, then, a QFP used to transform Fock states in frequency-bin encoding can be characterized according to either (i)~an open box viewpoint in which the system is modeled as a multiport interferometer [Eq.~\eqref{eq:J3}] or (ii)~a black box viewpoint in which the QFP is treated as an arbitrary quantum channel [Eq.~\eqref{eq:J7}]. Both approaches possess advantages and disadvantages, trading off between physical insight and generality. In the remainder of the paper, we discuss the findings of QFP experiments applying each technique.

\begin{figure}[tb!]
\centering
\includegraphics[width=3.5in]{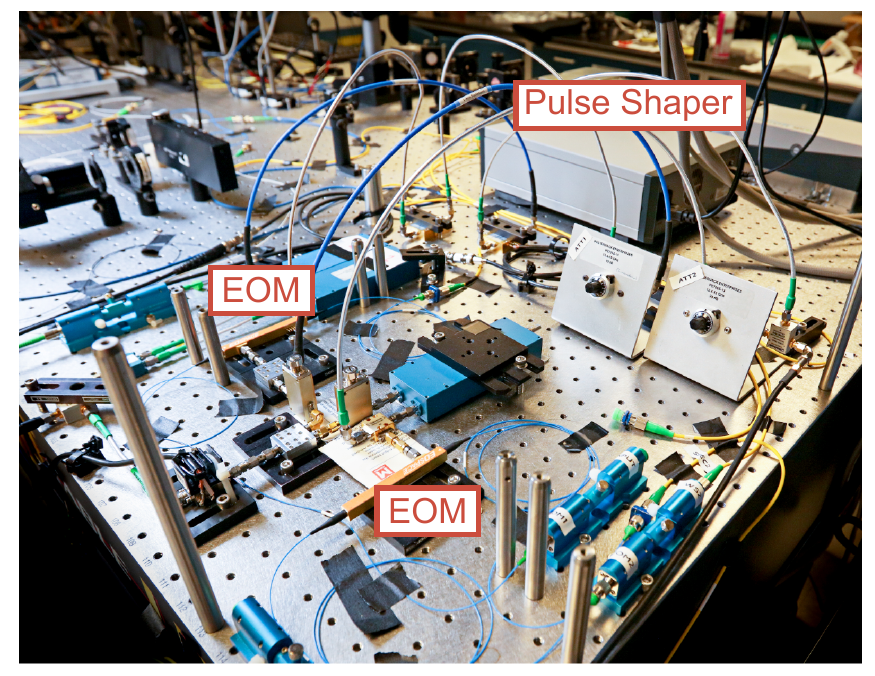}
\caption{Photograph of QFP used to demonstrate a frequency-bin tritter~\cite{Lu2018a}. The setup combines a commercial pulse shaper with two EOMs, each driven by a superposition of two microwave tones. The profusion of RF cables and optical fibers highlights the  mutual importance of electrical and optical facets of the QFP. Image credit: Jason Richards, Oak Ridge National Laboratory.}
\label{fig:2}
\end{figure}

\subsection{Experimental Setup}
\label{sec:ExpSetup}
Figure~\ref{fig:2} depicts an exemplary tabletop QFP consisting of two EOMs (EOSpace) and one pulse shaper (Finisar Waveshaper 1000S), which was experimentally leveraged to realize a balanced frequency tritter---a three-dimensional extension of the frequency beamsplitter~\cite{Lu2018a}. In addition to the optical components labeled in the setup, several RF components, including oscillator, frequency doubler, power splitters, amplifiers, attenuators, phase shifters, and cables, are employed to synthesize the drive signals for the EOMs. 
The spectral resolution of the pulse shaper (typically $\gtrsim$10 GHz) determines how tight the frequency bins can be packed, while the acceptance bandwidths of the RF components limit the maximum frequency separation. In this example, the mode spacing 
is chosen as $\Omega/2\pi=18.1$~GHz. Our frequency tritter design is based on both the fundamental tone $\Omega$ and the second harmonic $2\Omega$ in the EOM drive signals (specific solutions can be found in \cite{Lu2018a, Lu2022b}). To produce the necessary second harmonic, we split a portion of the 18.1~GHz sinewave from the oscillator and double its frequency to 36.2~GHz. We then use RF amplifiers, attenuators, and phase shifters to set the correct amplitude and phase for each RF tone before recombining them in an RF power coupler connected to the EOM. 

Applying spectral phase modulation via the commercial pulse shaper is comparatively simple---in the line-by-line shaping regime~\cite{Cundiff2010, Weiner2011}, we can program the target phase functions onto the pulse shaper without causing any distortions in the phase mask. Since EOMs naturally couple a single input frequency mode to many output modes, the pulse shaper must be able to address a larger number of modes than the input. For the tritter, the pulse shaper must shape a total of 12 frequency bins to reach $\mathcal{F}_W \geq 1-10^{-4}$, a small fraction of its total 40~nm (5~THz) bandwidth. In \cite{Lu2022b}, we show numerically that the same three-element (EOM/shaper/EOM) QFP can implement $d$-dimensional discrete Fourier transform (DFT) gates up to $d=10$ ($d=3$ for the frequency tritter here) with near-unity fidelity, when $d-1$ RF harmonics drive the EOMs and the pulse shaper applies phase shifts to a total of $\sim$4$d$ channels. This approach allows for the scalable construction of high-dimensional frequency-bin quantum operations that will be valuable for tasks such as entanglement verification~\cite{Spengler2012,Coles2017} and high-dimensional quantum key distribution~\cite{Sheridan2010,Islam2017b}.



\begin{figure*}[t!]
\centering
\includegraphics[width=6.5in]{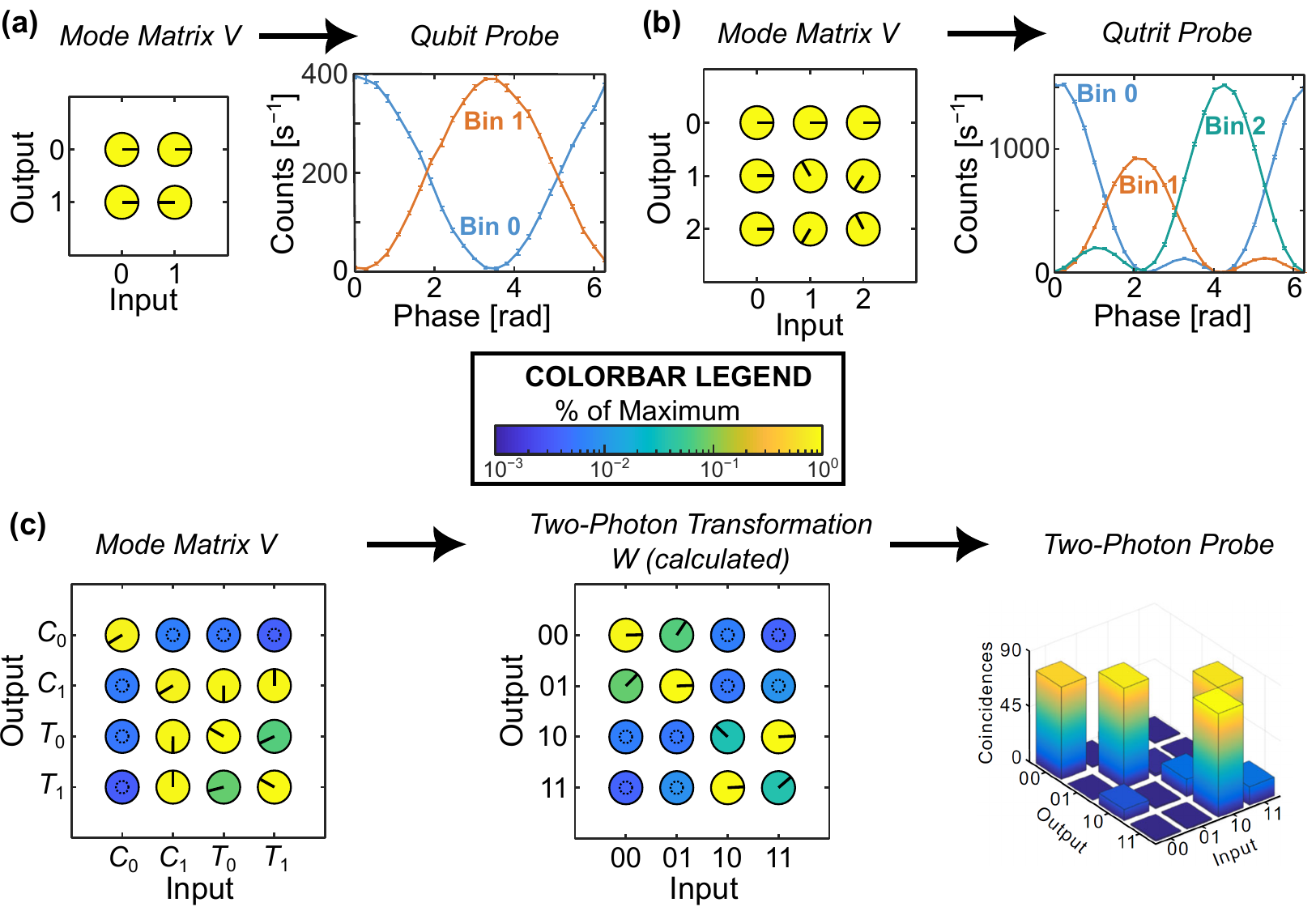}
\caption{Open box QFP characterization results. Each example shows a mode matrix $V$ recovered from bright coherent state probes, followed by photon counting results for quantum inputs. (a)~Frequency-bin Hadamard and (b)~tritter. For these two gates, the Hilbert space transformation $W$ and mode transformation $V$ are equal. (c)~Coincidence-basis CNOT. Here the two-photon transformation $W$ is calculated from the experimentally measured $V$, revealing a conditional bit flip as designed. For all matrix plots in (a--c), the color of each circle denotes the amplitude, and the radial line, the phase, of the relevant element; undefined phases (due to low amplitude) are depicted as dotted circles. Interferograms in (a,b) reprinted with permission from \cite{Lu2018a}. Images in (c) adapted from \cite{Lu2019a} under a \textcolor{blue}{\href{https://creativecommons.org/licenses/by/4.0/}{Creative Commons Attribution 4.0 International License}}.}
\label{fig:3}
\end{figure*}

\section{Open Box Characterization}
\label{sec:OpenBox}
\subsection{Pure Multiport}
Figure~\ref{fig:3} illustrates three representative operations realized on the QFP so far. In the first experimental example [Fig.~\ref{fig:3}(a)], the QFP was configured to operate as a Hadamard gate, also known as a balanced frequency beamsplitter~\cite{Lu2018a}. Mathematically the ideal transformation matrix can be written as 
\begin{equation}
\label{HadMat}
T_H = \frac{1}{\sqrt{2}} \begin{pmatrix} 1 & 1 \\ 1 & -1 \end{pmatrix}.
\end{equation}
To characterize the mode matrix $V$ via the open box approach, we first generated an  electro-optic frequency comb at the QFP mode spacing of 25~GHz and utilized an additional pulse shaper to adjust the amplitude and phase of each frequency mode. This enabled the preparation of either single-bin or superposition states. By probing the QFP with a single optical frequency mode ($\ket{1_{\omega_0} 0_{\omega_1}}$ or $\ket{0_{\omega_0} 1_{\omega_1}}$), the modulus of every matrix element in $V$ can be obtained by measuring the output optical spectra, as discussed in Sec.~\ref{sec:theory}. To determine the unknown phase terms in $V$, we probed the QFP with equal superpositions of two frequency modes ($\ket{1_{\omega_0}0_{\omega_1}} + e^{i\phi}\ket{0_{\omega_0}1_{\omega_1}}$), scanning $\phi \in [0,2\pi)$~\cite{footnote}. We extracted the power on specific modes from a series of optical spectra and observed the resulting interference patterns as a function of $\phi$. We then used these patterns to determine the unknown phase terms by fitting them to sinusoidal curves.

Figure~\ref{fig:3}(a) shows an experimentally obtained mode transformation for the beamsplitter, which corresponds to $\mathcal{P}_W = 0.9739$ and $\mathcal{F}_W = 0.9999$ when compared to the ideal $T_H$. To verify accuracy at the single-photon level, we attenuated the input state to $\sim$0.1 photons per detection gate window, and again scanned phase $\phi$. At each setting, a wavelength-selective switch (WSS) routed the output modes to an InGaAs single-photon detector. The resulting interference patterns [Qubit Probe in Fig.~\ref{fig:3}(a)] show the expected sinuosidal oscillations with fitted visibilities exceeding 97\%, which is primarily limited by the detector noise. 

By incorporating an additional harmonic to the EOM drives, as discussed in Sec.~\ref{sec:ExpSetup}, we experimentally reconfigured the QFP into a tritter in which the transformation matrix is a $3d$ DFT given by~\cite{Lu2018a, Lu2022b}
\begin{equation}
\label{DFTMax}
T_\mathrm{tr} = \frac{1}{\sqrt{3}} \begin{pmatrix} 1 & 1 & 1 \\ 1 & e^{2\pi i/3} & e^{4\pi i/3} \\ 1 & e^{4\pi i/3} & e^{2\pi i/3} \end{pmatrix}.
\end{equation}
We again applied the coherent-state-based gate characterization technique, now with a 18.1 GHz electro-optic frequency comb as the probe, and measured the $3\times3$ mode matrix $V$ as depicted in Fig~\ref{fig:3}(b). The moduli and phases match those of $T_\mathrm{tr}$ well, with an associated success probability of $\mathcal{P}_W = 0.9731$ and fidelity of $\mathcal{F}_W = 0.9992$. Again, to gain insight into the tritter for single-photon states, we attenuated a three-mode superposition state ($\ket{1_{\omega_0}0_{\omega_1}0_{\omega_2}} + e^{-i\phi}\ket{0_{\omega_0}1_{\omega_1}0_{\omega_2}} + e^{-2i\phi}\ket{0_{\omega_0}0_{\omega_1}1_{\omega_2}}$)  to the single-photon level and measured the counts in bins 0, 1 and 2 at the output. The interference patterns, now tracing a sum of two sines with respective peaks at $\phi\in\{0,2\pi/3, 4\pi/3\}$, again show excellent visibilities of over 97\%. The reduced flux in bin 1 is primarily due to the mismatch between the mode spacing (18.1~GHz) and the 12.5~GHz passbands on the WSS.

The previous two examples are one-photon gates, for which the the mode $V$ and Hilbert space $W$ transformations are identical; Eq.~\eqref{eq:J4} reduces to one term with a single factor. Two-qubit gates are also required for universal computing~\cite{Nielsen2000} which, following KLM~\cite{Knill2001}, can be realized 
via quantum interference and single-photon detection. Controlled-phase (CPhase) and controlled-NOT (CNOT) gates were first demonstrated in polarization/spatial encodings using a network of spatial beamsplitters~\cite{Hofmann2002,Ralph2002,OBrien2003}, where the gate success is postselected by coincidence events between the desired output ports. Mathematically, the ideal CNOT can be described in the computational basis as
\begin{equation}
\label{eCNOT}
T_\mathrm{CNOT} = \begin{pmatrix}
1 & 0 & 0 & 0 \\
0 & 1 & 0 & 0 \\
0 & 0 & 0 & 1 \\
0 & 0 & 1 & 0
\end{pmatrix},
\end{equation}
which flips the state of the target qubit if the control qubit is in state $|1\rangle$. Similarly, this coincidence-basis CNOT gate can be implemented in the spectral domain with a QFP ~\cite{Lu2019a}. 
Numerically, we identified a set of EOM/shaper modulation functions for the three-element QFP that results in theoretical fidelity $\mathcal{F}_W = 0.9999$ and success probability $\mathcal{P}_W = 0.0445$~\cite{Lu2019a}, close but slightly below the optimal success for a coincidence-basis CNOT of $\mathcal{P}_W=1/9$~\cite{Hofmann2002, Ralph2002} due to our experimentally limited circuit depth. In this solution, the indices for the logical bins are as follows: $(C_0, C_1, T_0, T_1) =  (0,6,7,8)$, where we use the notation $C_0~(T_0)$ and $C_1~(T_1)$ to denote the frequency bin index $m$ for the logical zero and one mode, respectively, of the control (target) qubit. Thus the four logical states can be defined in the Fock basis as $\ket{00}\equiv\ket{1_{C_0}0_{C_1}1_{T_0}0_{T_1}}$, $\ket{01}\equiv\ket{1_{C_0}0_{C_1}0_{T_0}1_{T_1}}$, $\ket{10}\equiv\ket{0_{C_0}1_{C_1}1_{T_0}0_{T_1}}$, and $\ket{11}\equiv\ket{0_{C_0}1_{C_1}0_{T_0}1_{T_1}}$.

Standard procedures for characterizing a photonic CNOT gate often involve preparing a set of input states that span the entire two-photon state space and then performing state tomography for each of these states at the output~\cite{OBrien2004}, which exceeded our resources (number of EOMs and pulse shapers) available at the time of the experiment. While the use of high-flux coherent states in the aforementioned characterization procedures cannot reveal the two-photon interference effects underlying the CNOT gate, it is still useful in characterizing the mode transformation $V$. Experimentally, we configured the QFP to realize the frequency-bin CNOT and measured the mode matrix $V$ shown in Fig.~\ref{fig:3}(c). As expected from the design, coupling between mode $C_0$ and $\{C_1, T_0, T_1\}$ was negligible, since $C_0$ is spectrally isolated from the target modes (to preserve the state of the target upon the presence of a photon in mode $C_0$). In contrast, bin $C_1$ is close to both target bins, allowing it to be strongly coupled to both $T_0$ and $T_1$ with equal strength. With the mode transformation $V$ at our disposal, the equivalent two-photon state transformation $W$ was computed from Eq.~\eqref{eq:J4}, which shows all four of the large elements of $W$ to be in-phase and in good agreement with the theory. We calculated an \emph{inferred} fidelity of $\mathcal{F}_W^{(\mathrm{inf})} =0.995$ and coincidence-basis success probability of $\mathcal{P}_W^{(\mathrm{inf})} =0.0460$---an indirect estimate that suggested our frequency-bin CNOT was operating correctly. 

To test our gate with truly quantum states, we generated a two-photon frequency comb---namely, entangled photons spanning discrete pairs of phase-coherent energy-matched comb lines---by pumping a periodically poled lithium niobate (PPLN) waveguide with a continuous-wave Ti:sapphire laser and filtering the broadband emission with an etalon to produce frequency bins. We then prepared all four computational-basis states by translating the pump frequency to four different values and using an additional shaper (prior to QFP) to filter out specific modes. After the gate operation, the output photons were frequency-demultiplexed 
and coincidences measured for 600~s for all 16 combinations of input/output states [plotted in Fig.~\ref{fig:3}(c)], showing a total of four distinct, significant peaks as well as a number of smaller bars. When an input photon is present in mode $C_0$, the quantum state is maintained, but when a photon is present in mode $C_1$, the target qubit is flipped. 

\subsection{Physically Motivated Noise Model}
\label{sec:CNOT}
Logical-basis measurements in the case of CNOT characterization, such as the one presented in Fig.~\ref{fig:3}(c), can only reveal the amplitude of the state transformation $W$; to access the phase information in $W$, measurements of superposition state inputs (for example, $(\ket{0}-\ket{1})\otimes\ket{0}$) are essential~\cite{OBrien2003}. However, we have found that by constructing a physical model that takes into account both single-detector events and coincidence counts, logical-basis measurements can provide more information than typically considered. In other words, by augmenting the multiport model in the open box viewpoint with physically motivated noise processes, it is possible to estimate the transformation $W$ without reference to the classical characterization in Fig.~\ref{fig:3}(c), but rather from the quantum data alone.

\begin{figure}[tb!]
\centering
\includegraphics[width=3.5in]{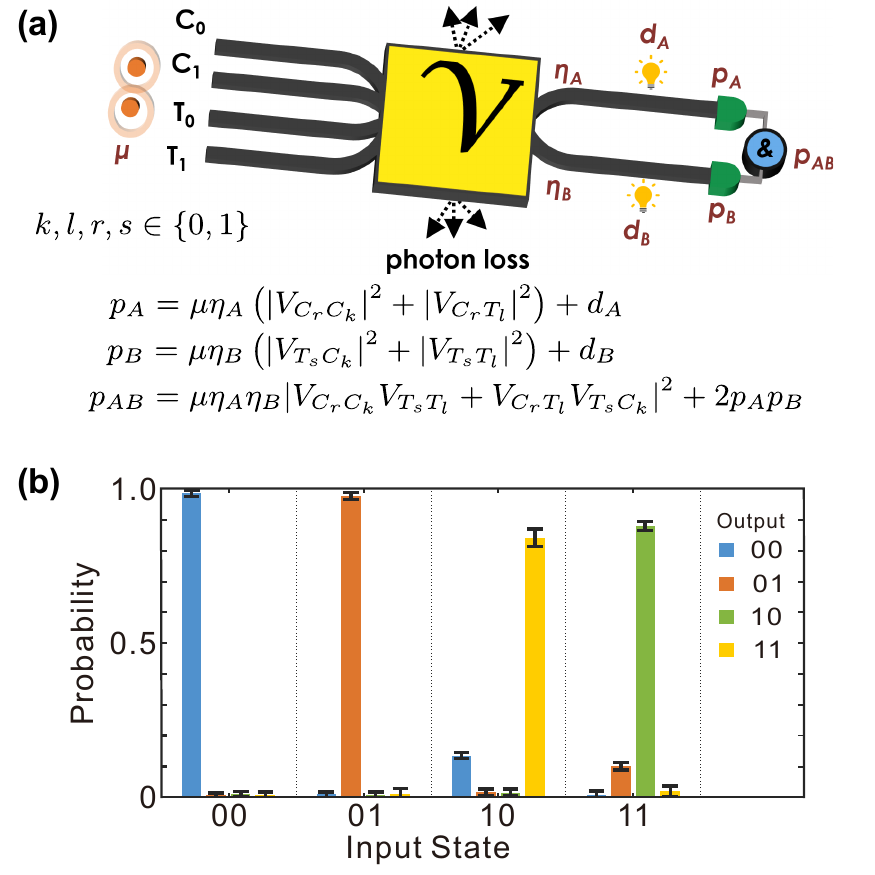}
\caption{Incorporating noise into the open box CNOT model. (a)~Contributions to single and coincidence detection probabilities including effects such as probabilistic pair generation ($\mu$), nonunity efficiency ($\eta_A, \eta_B$), background/dark counts ($d_A, d_B$), and uncorrelated events ($2p_Ap_B$). (b)~Transition probabilities recovered through Bayesian inference of this model with the two-photon results in Fig.~\ref{fig:3}(c). Panel (b) reprinted from \cite{Lu2019a} under a \textcolor{blue}{\href{https://creativecommons.org/licenses/by/4.0/}{Creative Commons Attribution 4.0 International License}}.}
\label{fig:4}
\end{figure}

Specifically, we considered a model [Fig.~\ref{fig:4}(a)] where logical input photon states 
$\ket{kl}$ ($k,l\in\{0,1\}$) are processed by a linear-optical frequency multiport fully described by a mode matrix $V$ (thus falling under the open box umbrella), and coincidences between output modes $C_r$ and $T_s$ ($r,s\in\{0,1\}$) are registered at detectors $A$ and $B$, respectively. We also accounted for nonideal effects in our model, including the pathway efficiency $\eta_A$ ($\eta_B$) from photon-pair generation to detector $A$ ($B$) and the dark count probability $d_A$ ($d_B$). For each set of $\{k, l, r, s\}$, we derived the marginal probabilities for clicks on detector $A$ or $B$, denoted as $p_A$ and $p_B$, respectively, defined within a detection frame $\tau$ ($\sim$1.5 ns in our case). The clicks can occur either due to the arrival of a photon from an entangled pair (with pair generation probability $\mu$) at the monitored output modes or due to random background or dark count noise ($d_A$ and $d_B$); under the assumption $\mu,\eta_A,\eta_B,d_A,d_B \ll 1$, we found the marginal probabilities to be simply the summation of pair and dark-count contributions (see \cite{Lu2019a} for detailed derivation).

The coincidence probability between detectors $A$ and $B$ ($p_{AB}$) also contains two terms---the correlated coincidences between two photons from the same entangled photon pair, and the accidental coincidences from two random clicks within the detection frame $\tau$. The latter term, represented by $2p_Ap_B$~\cite{Eckart1938, Pearson2010}, encompasses the noise coincidences from multipair emission or between a photon and dark count. Together, these equations show that the probability of detecting a single photon ($p_A$ and $p_B$) is determined solely by the magnitude of the elements in the $V$ matrix, whereas the coincidence probability ($p_{AB})$ also depends on the relative phase of these matrix elements. Consequentially, our goal was to devise a method to estimate the model parameters of interest $\beta = \{V,\mu,\eta_A,\eta_B\}$ from the measured datasets in Fig.~\ref{fig:3}(c).


The leading approach in modern quantum tomography, maximum likelihood estimation (MLE) is typically utilized to find a \emph{single} set of $\beta$ values that maximizes the likelihood function~\cite{Hradil1997, James2001}. However, we employed Bayesian inference to sample from a complete posterior distribution $P(\beta|\bm{\mathcal{D}})$, the probability density of the parameters $\beta$ given the observed data $\bm{\mathcal{D}}$. While more computationally intensive than MLE, Bayesian methods have the advantage of automatically quantifying uncertainties, and the Bayesian mean estimator attains the minimum squared error on average~\cite{Robert1999}. Although a detailed discussion of Bayesian techniques is beyond the scope of the present review, we direct the interested reader to the proposal and motivation for Bayesian QST from Blume-Kohout~\cite{Blume2010}, as well as several papers describing specific Bayesian tomography models and algorithms~\cite{DiGuglielmo2009, Seah2015, Granade2016, Mai2017, Williams2017, Lukens2020b, Chapman2022a, Lu2022a, Chapman2022b}.

For the considered CNOT experiment, we leveraged the Markov chain Monte Carlo (MCMC) algorithm known as slice sampling~\cite{Neal2003, Williams2017} to obtain Bayesian samples numerically, ultimately arriving at the fidelity estimate $\mathcal{F}_W = 0.91 \pm 0.01$. Figure~\ref{fig:4}(b) illustrates how this $\mathcal{F}_W$ translates into the output state probabilities for logical-basis inputs. Significantly, these results show the informative power of the open box QFP viewpoint: with Eq.~\eqref{eq:J3} combined with a reasonable noise model, truth table measurements alone [Fig.~\ref{fig:3}(c)] were in fact sufficient to estimate even the phases of the matrix $V$. 

\section{Black Box Characterization}
\label{sec:BlackBox}
In the previous section, we discussed experimental examples of the open box approach for characterizing the QFP. 
With the black box approach, Eqs.~(\ref{eq:J3},\ref{eq:J4}) are no longer assumed. Instead, one intentionally is given access to the input and output ports only, without any assumptions regarding the action of the QFP. By preparing a known input state and then performing measurements of the output state after the operation, the quantum process can be fully determined after repeating this procedure for a number of different input states in multiple bases.

In this section, we will focus on an exemplary class of QFP operations, namely, tunable frequency-bin beamsplitters. Previously in \cite{Lu2018b, Lu2020b}, we discovered that the three-element QFP can synthesize tunable frequency beamsplitters simply by adjusting the depth of the phase shift $\alpha$ imparted by the QFP shaper between frequency bins $\omega_0$ and $\omega_1$ while driving the EOMs with $\pi$-phase-shifted sinewaves with a fixed modulation index $\Theta$. 
Mathematically, the $2\times 2$ transformation matrix on the qubit Hilbert space is a function of the tunanble phase jump $\alpha$:
\begin{equation}
\label{tunableBS:1}
W =
\begin{pmatrix}
W_{00}(\alpha) & W_{01}(\alpha) \\
W_{10}(\alpha) & W_{11}(\alpha)
\end{pmatrix},
\end{equation}
where each of the matrix elements are
\begin{equation}
\label{tunableBS:2}
\centering
\begin{aligned}
W_{10}(\alpha) &= W_{01}(\alpha) = (1-e^{i\alpha}) \sum_{k=1}^\infty J_k(\Theta) J_{k-1}(\Theta) \\
W_{00}(\alpha) &=  J_0^2(\Theta) + (1+e^{i\alpha})\frac{1-J_0^2(\Theta)}{2} \\
W_{11}(\alpha) &=  e^{i\alpha} J_0^2(\Theta) + (1+e^{i\alpha})\frac{1-J_0^2(\Theta)}{2},
\end{aligned}
\end{equation}
with $J_k(\Theta)$ the Bessel function of the first kind.  
The identity operation is observed when $\alpha=0$, as the two $\pi$-phase-shifted sinewave cancel each other out. Furthermore, when $\alpha=\pi$, the reflectivity $\mathcal{R}\equiv |W_{01}|^2 = |W_{10}|^2$ and the transmissivity $\mathcal{T}\equiv |W_{00}|^2 = |W_{11}|^2$ are approximately equal; the elements $\{W_{00},W_{01},W_{10}\}$ are all real and positive, while $W_{11}$ is real and negative. This results in a Hadamard operation with theoretical fidelity $\mathcal{F}_W=1-10^{-7}$. Accordingly, the designed quantum process transforms any qubit input $\rho'$ into the output $\rho_T \propto W\rho'W^\dagger$. It is important to emphasize that this formula is not assumed in the characterization experiments below, but rather is used to compute the ideal output $\rho_T$ and process $\Phi_T$ against which experimental results are compared.

\subsection{Quantum State Tomography (QST)}
\label{sec:QST}
\begin{figure}[tb!]
\centering
\includegraphics[width=3.5in]{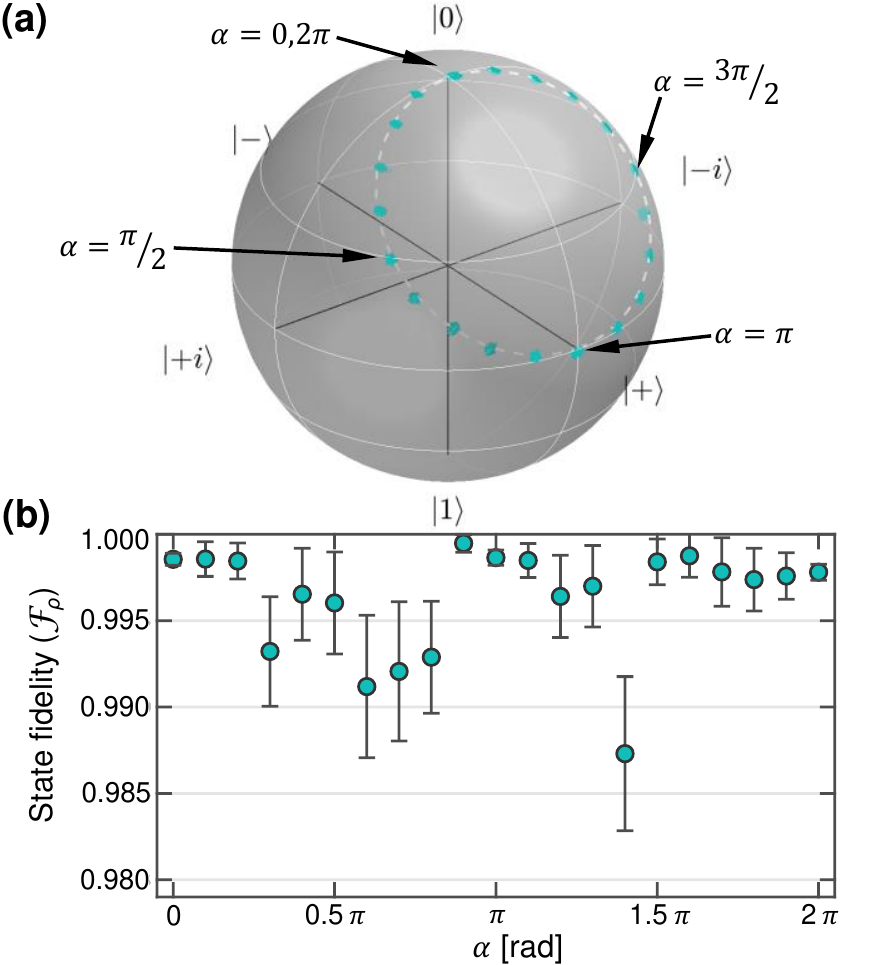}
\caption{Black box QST results for the tunable frequency beamsplitter. (a)~Measured density matrices $\rho$ for each beamsplitter setting $\alpha$ with fixed logical input state $\ket{0}$. Each ``cloud'' on the Bloch sphere consists of 1024 discrete samples obtained through Bayesian inference to quantify uncertainty. (b)~State fidelity $\mathcal{F}_\rho$ with respect to ideal output. Reprinted with permission from \cite{Lu2020b}.}
\label{fig:5}
\end{figure}

As the first experimental foray into black box QFP characterization, we investigated gate performance for converting a fixed input frequency-bin qubit to a targeted output state, both for arbitrary single-qubit unitaries and the tunable beamsplitter of interest here~\cite{Lu2020b}. We performed QST of the output photon and compared the recovered state $\rho$ against the target $\rho_T$ via Eq.~\eqref{eq:J8}. 
Specifially, we prepared the fixed single-frequency input $\ket{0}\equiv\ket{1_{\omega_0}0_{\omega_1}}$, 
using the tunable beamsplitter to convert this state, initially at the north pole of the Bloch sphere, to a family of states along a predefined trajectory [dashed line in Fig.~\ref{fig:5}(a)]; in this implementation, we sampled a total of 21 evenly spaced $\alpha\in[0,2\pi)$.

To perform QST on a single qubit, we measured projections onto the eigenvectors of the Pauli basis $\{X,Y,Z\}$---namely, $\ket{0}$, $\ket{1}\equiv\ket{0_{\omega_0}1_{\omega_1}}$, $\ket{\pm}\equiv\frac{1}{\sqrt{2}}(\ket{0}\pm\ket{1})$, and $\ket{\pm i}\equiv\frac{1}{\sqrt{2}}(\ket{0}\pm i\ket{1})$. These Pauli measurements are equivalent to applying specific quantum gates---$\mathbbm{1}$ for $Z$, $H$ for $X$, and $HS^\dagger$ for $Y$---followed by a measurement in the computational basis, where $\mathbbm{1}$ is the identity operation, $H$ is the Hadamard, and $S=\begin{psmallmatrix} 1 & 0 \\ 0 & i \end{psmallmatrix}$. For the $H$ gate required in both $X$ and $Y$ measurements, we opted for an inherently probabilistic single-EOM version~\cite{Imany2018,Seshadri2022} that is simpler than the QFP Hadamard~\cite{Lu2018a}. 
This method attains a theoretical fidelity of $\mathcal{F}_W = 1$ and success of $\mathcal{P}_W \approx 0.6$). The $S^\dagger$ gate amounts to a phase shift between two frequency modes, which can be realized by using an extra pulse shaper, which also blocks any residual photons outside of the single-qubit space that may occur due to nonunity success of the QFP operations. 
As the final step for all three Pauli measurements, we demultiplex the photons by color with a WSS and record the counts over 1~s with superconducting nanowire single-photon detectors (SNSPDs), obtaining the full dataset of counts for all basis states. 

As with the CNOT in Sec.~\ref{sec:CNOT}, we employed Bayesian inference to estimate the quantum output state for each beamsplitter setting, this time leveraging preconditioned Crank--Nicolson MCMC sampling~\cite{Cotter2013} after the approach of \cite{Lukens2020b}. Figure~\ref{fig:5}(a) plots the resulting states, where each spot consists of a cluster of 1024 samples from the Bayesian posterior distribution. The good agreement with the expected trajectory is confirmed quantitatively by the fidelity calculations in Fig.~\ref{fig:5}(b), whose means all exceed 0.985.

\subsection{Quantum Process Tomography (QPT)}
The previous QST results above are truly ``black box'' in that they make no reference to the inner workings of the QFP. Yet strictly speaking, they confirm ideal functioning of the tunable beamsplitter for a \emph{single} input only---namely $\rho'=\ket{0}\bra{0}$. 
To fully characterize a quantum process, it is necessary to repeat the aforementioned procedure with a variety of input states, including frequency-bin superpositions that necessitate an additional EOM and pulse shaper beyond what was available to us during the original QST experiments~\cite{Lu2020b}. With a sufficient number of devices now in hand, however, we proceed with full QPT of the tunable beamsplitter---the first QPT experiment of a frequency-bin quantum gate.

\begin{figure}[tb!]
\centering
\includegraphics[width=3.5in]{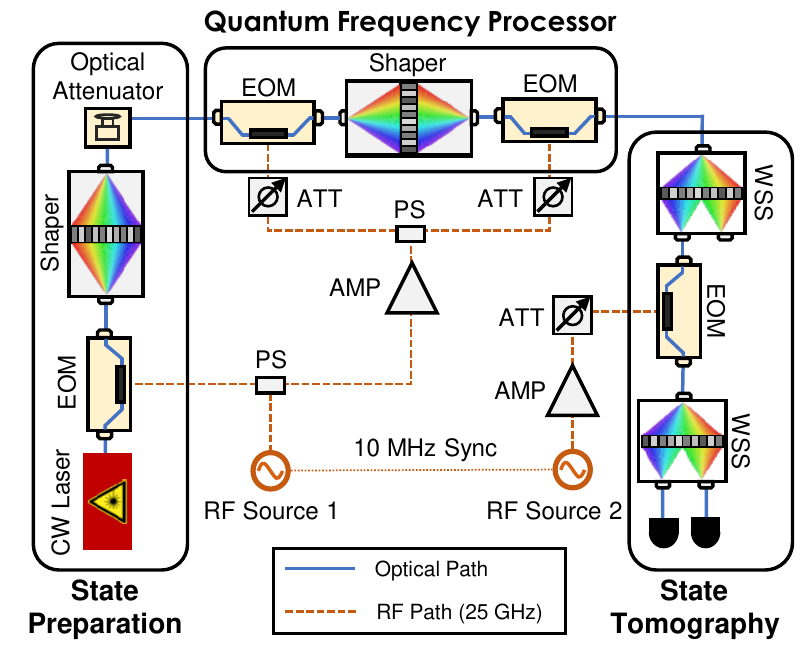}
\caption{Experimental QPT setup for tunable frequency-bin beamsplitter. Single-photon-level inputs traverse the QFP and proceed through a WSS (to block bins outside of the computational space), an EOM (to interfere frequency bins), and a final WSS to separate computational bins for detection. All manual RF phase shifters have been supplanted in favor of electrically controlled options, both pulse-shaper-induced delays and direct programming of RF Source 2's phase with respect to RF Source 1. RF component labels: AMP (amplifier), ATT (attenuator), PS (power splitter).}
\label{fig:6}
\end{figure}

In standard QPT, a total of $d^2$ states permit an informationally complete basis for an arbitrary $d$-dimensional quantum process~\cite{Chuang1997, Nielsen2000}. A typical approach for single-qubit QPT ($d=2$) is to prepare $\{\ket{0}, \ket{1}, \ket{\pm}\}$ and perform QST on the output state for each.
In our frequency-bin QPT experiment, we consider an overcomplete set of input states, specifically the Pauli eigenstates defined in Sec.~\ref{sec:QST}: $\{\ket{0}, \ket{1}, \ket{\pm}, \ket{\pm i}\}$.

Figure~\ref{fig:6} provides a schematic of the experimental setup. The presence of four EOMs implies the need to precisely set three RF delays, and in contrast to previous experiments with manual phase shifters, we here control these delays electronically: two are tuned by applying a linear phase on the first two pulse shapers, and the third is adjusted by setting the phase of RF Source 2 relative to RF Source 1. To prepare the input states we again utilize amplitude and phase filtering of a 25~GHz electro-optic frequency comb with a state preparation pulse shaper. Prior to the QFP, we reduce the input photon flux to approximately 10$^6$ counts per second (roughly one-tenth of the detector saturation level) with an optical attenuator, mimicking the case of having true single photons with similar flux at the input.

The central QFP is configured to realize a tunable frequency beamsplitter, in this case, with reflectivities $\mathcal{R}\in\{0,0.123,0.373,0.5\}$ corresponding to $\alpha\in\{0,\pi/3, 2\pi/3,\pi\}$ in Eq.~\eqref{tunableBS:2}. We continue the single-EOM approach for tomographic projections, but with two slight modifications: (i)~a dual WSS (Finisar) is utilized to block photons outside the computational space after the QFP and frequency-demultiplex the photons before the SNSPDs, and (ii)~the $S^\dagger$ operation is realized through precise timing adjustments on the State Tomography EOM, 
leveraging the equivalence between linear spectral phase and delay. Finally, we count photons at the output with 1~s integration time, and do not subtract dark counts. The full dataset for each QFP setting thus contains 36 numbers: counts for all six output projections, repeated for each of the six input states.

\begin{figure*}[tb!]
\centering
\includegraphics[width=6.5in]{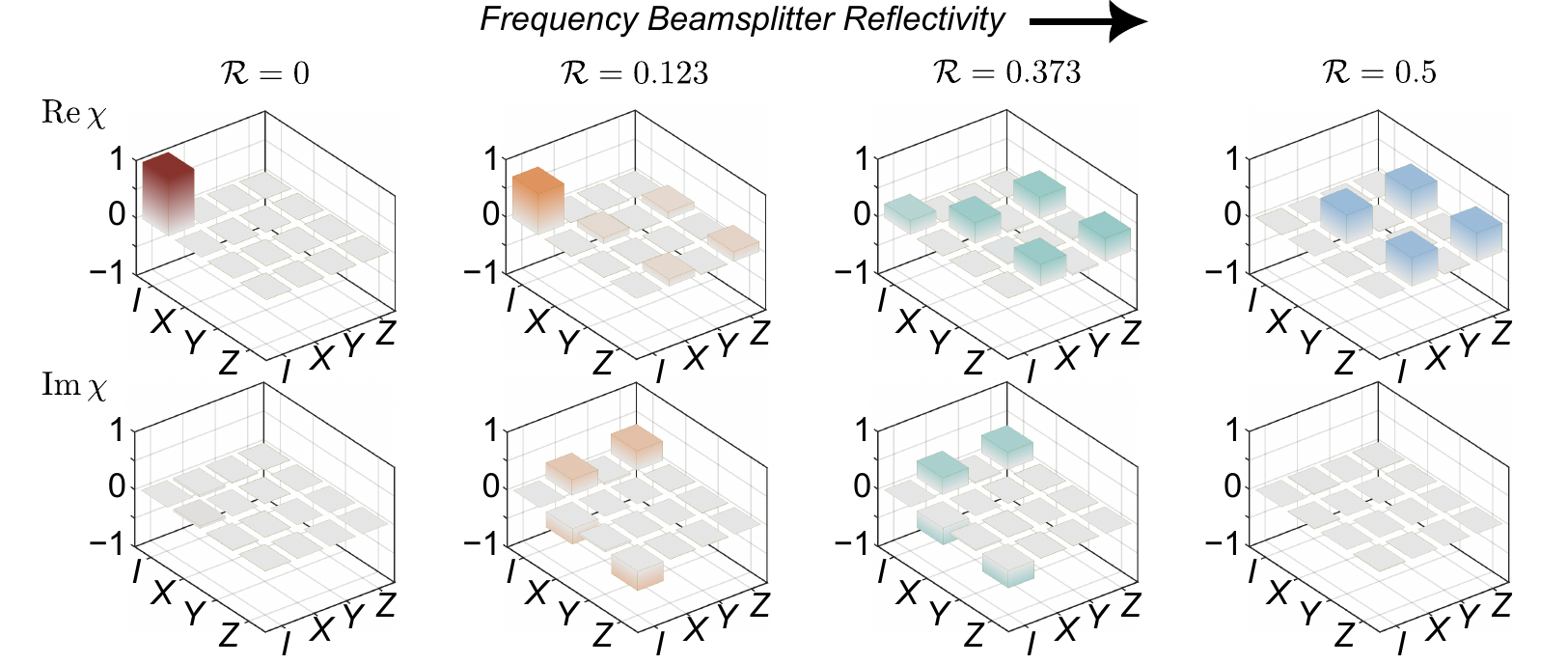}
\caption{Experimental QPT results for tunable frequency-bin beasmplitter with varying reflectivity $\mathcal{R}$. Plotted is the Bayesian mean process matrix $\chi$ in the Pauli basis. As $\mathcal{R}$ increases, the initially dominant identity component reduces, while the ideal $X$ and $Z$ contributions for the Hadamard rise. Process fidelities $\mathcal{F}_\Phi$ with the respect to the designed beasmplitter are (left to right): $0.9925\pm0.0002$, $0.9775\pm0.009$, $0.9878\pm0.0009$, and $0.9946\pm0.0002$.
}
\label{fig:7}
\end{figure*}

For inference, we once again subscribe to a Bayesian paradigm, following  the QPT procedure introduced in \cite{Chapman2022b} which pools all 36 numbers into a single likelihood and employs a uniform prior based on a recently proposed Kraus operator parametrization~\cite{Kukulski2021}. The real and imaginary parts of the Bayesian mean process matrices $\chi$ follow in Fig.~\ref{fig:7}, expressed in the Pauli basis for convenience. The process for $\mathcal{R}=0$ ($\alpha=0$) matches the identity as expected, with process fidelity $\mathcal{F}_\Phi = 0.9925\pm0.0002$, while that for $\mathcal{R}=0.5$ ($\alpha=\pi$) aligns with the Hadamard with $\mathcal{F}_\Phi = 0.9946\pm0.0002$. 
The intermediate cases reveal a clear transition between the two extremes of a single nonzero component for the identity ($\mathcal{R}=0$) and four equal components in the $X/Z$ subspace ($\mathcal{R}=0.5$), matching theoretical expectations with high fidelity: $\mathcal{F}_\Phi = 0.9775\pm0.0009$ for $\mathcal{R}=0.123$ and $\mathcal{F}_\Phi = 0.9878\pm0.0009$ for $\mathcal{R}=0.373$.

\section{Discussion}
\label{sec:discussion}
The open box and black box approaches for QFP characterization trade off advantages in complementary ways, so the method of choice for a given experiment can vary. From a design perspective, the QFP is the realization of some desired mode transformation matrix through sequential temporal and spectral phase modulation. Because this multiport interferometer applies to any optical input, open box approaches are able to leverage bright coherent states as probes, enabling extremely precise characterization without the additional complications of quantum probes sensitive to statistical noise and background light. Accordingly, open box QFP characterization provides the most accurate insight into the QFP's inner workings, and therefore is well suited to the engineering phase of quantum gate synthesis when the adjustable parameters of each device are tuned---e.g., EOM voltages and pulse shaper phases.

On the other hand, the noise effects bypassed by the open box formalism cannot be ignored in the quantum domain. Thus, after the QFP is tuned and ready for specific quantum information processing tasks, more general black box characterization approaches should be preferred, ideally performed at whatever flux and statistical properties will be used in the application of interest. QPT incorporates all nonidealities present in the system, consequently painting the most comprehensive picture of the final QFP operation.

Over the short history of the QFP, open box characterization has proven highly successful, which can be attributed to the fact that experimental implementations with commercially available EOMs and pulse shapers have been near-ideal, to an extent surpassing even our own original expectations. But this situation will likely change significantly as the QFP enters new application regimes. For example, fully on-chip QFPs have the potential for lower loss, tighter bin spacings, and reduced cost compared to tabletop versions, but on-chip components tend to deviate from the ideal line-by-line pulse shaper and linear EOM models. As explored in \cite{SiPhQFP}, microring-based pulse shapers lead to matrix transformations $V$ that vary depending on spectral location within a bin, while standard silicon phase modulators suffer from phase-dependent loss. Similarly, time synchronization of QFPs across multiple sites---required for applying the QFP to distributed quantum networking---will experience at least some level of jitter that, depending on its severity, could cause the transformation $V$ to drift noticeably in time. 
It is ultimately the black box viewpoint that possesses sufficient generality to integrate all such impairments into a single channel description
, which in our opinion points to the increasing importance of QST and QPT in next-generation QFP experiments.

Finally, although we have focused on the QFP due to the variety of experimental characterization demonstrations so far, the techniques described in this paper readily apply to other approaches for frequency-bin quantum operations. For example, frequency beamsplitters~\cite{Raymer2010} have been realized in $\chi^{(2)}$~\cite{Kobayashi2016, Kobayashi2017} and $\chi^{(3)}$~\cite{Clemmen2016, Joshi2020} nonlinear materials mediated by classical pump pulses, as well as in integrated microrings with coupling controlled by electro-optic modulation~\cite{Hu2021}. While compatible with single-photon signals, these experimental frequency-bin gates have yet to be fully characterized in either the open box multiport or black box quantum process pictures. In the future, it would therefore prove interesting to invoke the techniques described here for non-QFP-based frequency-bin operations, facilitating a more comprehensive and informative understanding of the ever-expanding toolkit for frequency-bin quantum information.

\begin{acknowledgments}
We thank P. Imany, D.~E. Leaird, P. Lougovski, O.~D. Odele, E.~M. Simmerman, and B.~P. Williams for contributions to past publications summarized in this article~\cite{Lu2018a, Lu2019a, Lu2020b}. We thank M. Alshowkan for setting up the time tagging electronics. This research was performed in part at Oak Ridge National Laboratory, managed by UT-Battelle, LLC, for the U.S.~Department of Energy under contract no.~DE-AC05-00OR22725.
\end{acknowledgments}


\end{document}